\documentclass[11pt]{article}

\usepackage{amsmath, amssymb, wasysym, slashed, multirow,hyperref}
\usepackage{pdfpages}
\usepackage{cite}
\usepackage{times}

\newenvironment{sciabstract}{%
\begin{quote} }
{\end{quote}}

\newcounter{lastnote}

\title{Resurgence of the Renormalization Group Equation}

\author
{Jahmall Bersini,$^{1\dagger}$ Alessio Maiezza,$^{1\ast}$ Juan Carlos Vasquez$^{2\ddagger}$\\
\\
\normalsize{$^{1}$Ruder Bo\v skovi\'c Institute, Bijeni\v cka cesta 54, 10000, Zagreb, Croatia,}\\ \\
\normalsize{$^{2}$Amherst Center for Fundamental Interactions, Department of Physics,} \\ \\
 \normalsize{$$ University of Massachusetts, Amherst, MA 01003, USA. }\\
\\
\small{ E-mail: Jahmall.Bersini@irb.hr$^{\dagger}$,Alessio.Maiezza@irb.hr$^{\ast}$,jvasquezcarm@umass.edu$^{\ddagger}$}}

\usepackage{fancyhdr}
\fancypagestyle{plain}{%
	\fancyhead[R]{ACFI-T19-13}
	
}

\date{}

\begin{document}


\baselineskip16pt 


\maketitle

\begin{sciabstract}
We show how the renormalons  emerge from the renormalization group equation with \textit{a priori} no reference to any Feynman diagrams. The proof is rather given by recasting  the renormalization group equation as a resurgent equation studied in the mathematical literature, which describes a function with an infinite number of  singularities in the positive axis of the Borel plane. Consistency requires a one-to-one correspondence between the existence of such kind of equation and the actual (generalized) Borel resummation of the renormalons through a one-parameter transseries. Our finding suggests how non-perturbative contributions can affect the running couplings. We also discuss these concepts within the context of gauge theories, making use of the large number of flavor expansion.
\end{sciabstract}

\section{Introduction}

 By consistently removing the divergences of the loop integrals coming from the perturbative expansions, the renormalization procedure is a fundamental tool that makes sense of Quantum Field Theory (QFT). Nevertheless, the \emph{perturbative} renormalization prescriptions are somehow incomplete and partial because the series obtained are asymptotic and sometimes non-Borel summable due to instantons and renormalons~\cite{tHooft:1977xjm}, being the latter divergences in the Borel plane appearing during the renormalization procedure. Recently in Ref.~\cite{Maiezza:2019dht}, renormalons have been resummed in a generalized sense within the framework of the \emph{analyzable functions}~\cite{Costin1995,costin1998,Costin2008}, leaving only one unknown constant. As defined in Ref.~\cite{costin1998}, analyzable functions have a unique associated transseries which are Borel summable after a finite number of transformations.

Following the definition in Ref.~\cite{Costin2008}, the process of obtaining the actual function using the formal (divergent) transseries is called the ``\emph{synthesis}'' of the expression that one is attempting to resum. This is the result obtained in Ref.~\cite{Maiezza:2019dht}, where the synthesis has been performed
on renormalons appearing in the scalar field model and $SU(3)$ gauge model with a large number of flavors. We should stress that the synthesis must be in a \emph{one-to-one} correspondence with a specific non-linear ordinary differential equation (ODE)~\cite{Costin2008}. The study of the properties of the transseries from this unique ODE is called the ``\emph{analysis}'' of the ODE that one is interested in. In Ref.~\cite{Maiezza:2019dht}, it has been pointed out that the ODE for renormalons is indeed the Renormalization Group Equation (RGE) or the Callan-Symanzik equation~\footnote{These two equations are often identified in the literature (see for instance Ref.~\cite{Peskin:1995ev})) as the same equation, although they differ for considering or not the non-homogeneous term. In this article, we also use the two names regardless of this distinction.}.

Motivated by the result of the resummed renormalons together with the correspondence between synthesis and analysis, the scope of the present article is to let the renormalons emerge formally from the RGE, with no reference to explicit loop calculations. More precisely, we shall prove that an infinite number of singularities of the Borel transform of the Green function emerges directly from the homogeneous RGE. They have the usual properties of the renormalons. This result closes the two ways relation between the synthesis and the analysis for renormalons.

It is worth recalling that, although there is compelling evidence for the existence of renormalons~\cite{Bauer:2011ws}, a generic proof is still missing. We believe that ours may be the long-sought proof of their existence in a generic QFT and it is rooted in one of the most general principles of QFT, namely the RGE. This should settle the dust and eliminates any concern sometimes risen in the literature about their actual existence~\cite{Suslov:2005zi}.

We also provide illustrating examples within the well-known large $N_f$ expansion. The reason is twofold: large $N_f$ naturally suppresses the instantons~\cite{PhysRevD.15.1655,KOPLIK1977109}, then emphasizing the renormalon ambiguities; it leads to ``exact'' results in the leading $1/N_f$ expansion that allows obtaining insights on renormalons and their impact on QFT\footnote{For previous studies on the application of resurgence ideas in field theory see Refs.\cite{Dunne:2012ae,PhysRevLett.112.021601,Bellon:2016med}. }.

The paper is organized as follows. In Sec.~\ref{SEC:Setup}, we introduce the issues one encounters in the Borel representation of QFT; in Sec.~\ref{SEC:MAIN}, we show how renormalons emerge from the Callan-Symanzik equation. This is the central point of the article. In Sec.~\ref{SEC:largeN} and Sec.~\ref{SEC:cha}, we make contact with gauge theories illustrating the appearance of renormalons in specific examples. Finally, the work is equipped with three appendices: ~\ref{rec} analyze the nonperturbative contribution to a recursive equation which follows from the RGE; ~\ref{Appendix:nonliearODE} recalls properties of ODEs and shows some explicit manipulations; ~\ref{App:OSBETA} contains several details on large $N_f$ expansion within the on-shell scheme in quantum electrodynamics (QED).

\section{Perturbative vs nonperturbative QFT}\label{SEC:Setup}

Let us start stressing the importance of the Borel-Laplace integral representation of QFT. In principle, this is a non-perturbative tool that has the advantage of being built from perturbation theory. A parallel remark may be also made with the usual integral representation, i.e. the Feynman path integral.

\paragraph{From the Feynman path integral to perturbation theory}
In QFT one can write down all the fundamental Green functions in the schematic form of a path integral
\begin{equation}
G(g) = \int \mathcal{D}\phi e^{iS(g)}\,,
\end{equation}
where $S(g)$ is the action functional that depends on the coupling constant $g$. Although it is a powerful tool for generic proofs in QFT, explicit calculations usually require an expansion of $S(g)$ for small coupling $g$ (perturbation theory).
Such an expansion is mathematically ill-defined because one interchanges the summation of the series expansion with the functional integral. This operation is indeed legitimate only when the conditions of monotone convergence (Levi's theorem) are fulfilled. This does not hold in QFT and $S$-matrix expansion, leading to the well-known divergence of the perturbative series, pointed out for the first time by Dyson~\cite{PhysRev.85.631}.

\paragraph{The Borel-Laplace representation}
The Borel-Laplace representation may be considered a complementary representation, capable to provide results valid for finite (and potentially large) values of the coupling:
\begin{equation}\label{BLrepr}
G(g) = \int_0^{\infty} d z e^{-z/g} B(z)\,,
\end{equation}
where $B(z)$ is called the Borel transform of $G(g)$. Suppose you can reconstruct $B(z)$ from perturbation theory, then you have calculated a nonperturbative correlation function. This reconstruction may be approximately done through the well-known Borel-Pad\'e approach, or via more recent and sophisticated method involving
Meijer G-functions proposed in Ref.~\cite{Mera:2018qte} and employed in QFT in Ref.~\cite{Antipin:2018asc}.

\paragraph{Limitations of the Borel-Laplace representation}

\begin{enumerate}

\item \textbf{Borel transform singularities.} It is well known that poles anywhere in the complex plane of the Borel variable $z$ limit the radius of convergence and it is also the reason for the emergence of the $n!$ growth of the perturbation theory at a given order $g^n$.
When the Borel transform $B(z)$ has poles in the positive real axis, the Laplace integral becomes ambiguous and the resummation procedure fails to preserve the reality of the perturbative expansion. This is exactly what happens with the renormalons in QFT, either at the ultraviolet (UV) or the infrared (IR) limit, where renormalons appear during the process of renormalization\footnote{It is worth recalling that renormalons can emerge even in quantum mechanics whilst there is the necessity of renormalization, as recently shown in Ref.~\cite{Pazarbasi:2019web}.}.
The problem becomes even more severe in a model with more couplings and fields coupled with each other because Borel singularities move closer to the origin~\cite{Maiezza:2018ags}.

Fortunately, there exists a generalized Borel-Laplace resummation procedure developed by Costin~\cite{Costin1995,costin1998,Costin2008}. It can resum the perturbative series when $B(z)$ has an infinite number of poles in the positive real axis, preserving the reality of the perturbative series. This resummation procedure has been recently applied in QFT in connection with RGE~\cite{Maiezza:2019dht}, in which a non-perturbative mass is generated  for strongly coupled QFTs, as previously found in Refs.~\cite{Bellon:2014zxa,Bellon:2016mje}.

\item \textbf{Super-exponential growth.} Unfortunately, this is not the end of the story, since as originally argued by 't Hooft in Ref.~\cite{tHooft:1977xjm}, even if the Borel-Laplace integral is free of ambiguities, the Green functions still have singularities in the $g$-complex plane when
\begin{equation}
\frac{1}{g} = C + \frac{\beta_1}{2}(2n+1) \pi i \,,
\end{equation}
where $C$ is an arbitrary real constant. These singularities mean that
\begin{equation}
G(g) = \int^{\infty}_0 d z B(z) e^{[-c+\frac{\beta_1}{2}(2n+1) \pi i ]z}
\end{equation}
must diverge and therefore $B(z)$ must grow faster than any exponential of $z$ when $z\rightarrow \infty$. A pioneering way to deal with this problem has been proposed in Ref.~\cite{tHooft:1977xjm}, using a so-called ``second Borel procedure". Perhaps a general and systematic approach to the super-exponential behavior should be within the context of acceleration theory~\cite{Ecalle1993}-- see  Ref.~\cite{Bellon:2018lwy} for  discussions of these ideas in field theory. While this problem is intriguing and worth to be at least recalled here, it is beyond the scope of this work. From here on, we stick to the case when the Borel-Laplace representation of $G(g)$ is finite for $g<g_0$, with $g_0$ some real number. In other words, we assume that $G(g)$ is perturbatively described by a Gevrey-1 type formal power series.

\end{enumerate}

\section{From the renormalization group equation to ultraviolet renormalons }\label{SEC:MAIN}

In perturbation theory the two-point correlation function may be written as
\begin{equation}\label{G_PT}
G \sim \sum_i^{\infty} c_i(L) g^i\,,
\end{equation}
where $L=\ln(-q^2/\mu^2)$ is the scale dependence and the symbol "$\sim$" indicates that an asymptotic series for the Green function is used.
Eq.~\eqref{G_PT} can be rearranged in a convenient form (see for example Ref.~\cite{Peskin:1995ev}), which shall be the starting point for our manipulations:
\begin{equation}\label{G_PT2}
G \sim\sum_i^{\infty} \gamma_i(g) L^i\,.
\end{equation}
Order by order in perturbation theory, one can normalize $\gamma_0=1$ by a proper choice of counterterms. This can be achieved, for instance, by absorbing the finite part of the Green function at any perturbative order in the counterterms. Hence $G$ can be written in powers of $L$ coming from the divergent part of loop integrals. Plugging Eq.~\eqref{G_PT2} in the renormalization group equation
\begin{equation}\label{CS}
[- \partial_L+\beta(g) \partial_g -\gamma]\,G(L,g)=0\,,
\end{equation}
where $\beta(g) = \frac{dg}{d\log\mu^2}$ and $\gamma$ is the anomalous dimension, one gets a recursive equation in $\gamma_i$~\cite{Yeats:2008zy,Klaczynski:2013fca} which start in $\gamma_1=\gamma$ (see below in~\ref{rec})
\begin{equation}\label{rec_building}
(k+1) \gamma_{k+1}(g) = \beta(g) \frac{\partial \gamma_k}{\partial g}- \gamma(g)\gamma_k(g).
\end{equation}

This is not the end of the story since it is fundamental to realize that the ``order by order'' procedure used to write the expressions in Eqs.~\eqref{G_PT} and~\eqref{G_PT2} is mathematically ill-defined as $i\rightarrow \infty$, because these series are asymptotic. To rewrite Eq.~\eqref{G_PT2} with an exact equal sign, we may rewrite $G$ with a generic
function $R(g)$ to take into account possible non-perturbative contributions. One therefore has
\begin{equation}\label{G_true}
G(L,g) = 1-\sum_{i=1}^{\infty} \gamma_i(g) L^i + R(g) \,,
\end{equation}
where we have redefined for convenience the expression in the summation with a minus sign. As it shall be clear below in~\ref{rec}, considering $R$ has also an impact on the anomalous dimension $\gamma$. Notice that we do not assume anything for the function $R$, except that it is a nonperturbative object. This already suggests that $R$ should be related to some $n!$-growing of the perturbative expansion, but we do not need such a hypothesis. On the contrary, the properties of $R$ and its Borel transform emerge from our analysis.

We anticipate that the central result of this work is to show that $R(g)$ and in particular $\gamma$ obey a particular ODE, coming from the RGE for $G$, upon which Costin's resurgence theory is built. This equation must be in the form
\begin{equation}\label{mainEq}
R'(x)=- \frac{R(x)}{\beta_1}+ F_R(R,x)\,,
\end{equation}
or in term of the $\gamma$ function
\begin{equation}\label{mainEq2}
\gamma'(x)=- \frac{\gamma(x)}{\beta_1}+ F_\gamma(\gamma,x)\,,
\end{equation}
being $x=1/g$, $\beta_1$ the one-loop coefficient of the $\beta-$function $\beta(g)= \beta_1g^2+\beta_2 g^4...$ and $F_A=\mathcal{O}\left( A^2|x^{-1}|A x^{-1} \right)$ with $A=R(x),\gamma(x)$. Note that for the resurgent analysis below, only the perturbative expansion of $\beta(g)$ function is needed.
Because of the nonlinearity in $\gamma$ inside $F_\gamma(\gamma,x)$, the Borel transform of $\gamma(g)$ has a Stokes
line with infinite singularities at $n/\beta_1$ with $n=1,2,3,...$. These can be identified with the renormalons. Thus, Eq.~\eqref{mainEq2} lets the renormalons emerge
with no reference to any Feynman loop, e.g. the 't Hooft skeleton diagram. This central equation is proved in the next subsection, while specific realizations and details are shown in~\ref{Appendix:nonliearODE} (see the review in Ref.~\cite{Dorigoni:2014hea}).

\subsection{The resurgent non-linear ordinary differential equation}

In this section we elaborate on the proof of Eq.~\eqref{mainEq2}. First, consider a non-asymptotically free model, namely $\beta_1>0$. For our proposes, it is sufficient to expand the first leading terms of Eq.~\eqref{G_true}
\begin{equation}\label{G_true_leading}
G \simeq 1 -\gamma_1(g) L + R(g) \,,
\end{equation}
with $\gamma_1$ perturbative. Eq.~\eqref{G_true_leading} is the usual textbook expression, being $R(g)$ the finite part. The only new ingredient that we introduce is the assumption that $R(g)$ is nonanalytic (nonperturbative) - besides this, we do not require any specific feature for either $R$ or its Borel transform. Note also that $R$ cannot be removed from the Green function by a particular choice of the energy scale ($L=0$).

Plugging $G$ in Eq.~\eqref{CS} and collecting the terms of the order $L^0$, one gets\footnote{The non-perturbative piece $R(g)$ affects higher order terms in $L$ (see~\ref{rec}).}:
\begin{equation}\label{ODE0}
(R(g)+1)\gamma(g)=R(g)'\beta(g)+\gamma_1(g)\,.
\end{equation}
If $R=0$ then $\gamma=\gamma_1$ and $G$ can be perturbatively built recursively~\cite{Yeats:2008zy,Klaczynski:2013fca} using Eq.~\eqref{rec_building}; vice versa,
if $\gamma=\gamma_1$, $R$ is analytic as dictated by Eq.~\eqref{ODE0}, namely perturbative in $g$ in contradiction with the hypothesis that it is a nonperturbative correction to $G$. Thus it can be just reabsorbed as usual in $\gamma_0$ in Eq.~\eqref{G_PT2}, and this is effectively equivalent to redefine it $R=0$. Therefore, from the double-implication $\gamma=\gamma_1 \,\,\, \Leftrightarrow \,\,\, R=0$, it follows that there must exist a function $M(R,g)$ mapping the nonperturbative effects in $\gamma$ with the ones in $G$ such that
\begin{equation}\label{gen_gamma}
\gamma=\gamma_1+M(R(g),g), \,\,\,\,\,\,\,\,\, M(0,g)=0\,.
\end{equation}
For the resurgent result in Eqs.~\eqref{mainEq} and~\eqref{mainEq2}, the only relevant expansion
is in the form of a Taylor series $M(R,g)=q R(g)+\frac{1}{2}(r R(g)^2+2s g R(g)) + \mathcal{O}(R^3|g^2)$, being $q,r,s$ arbitrary complex constants. Using this expansion in Eq.~\eqref{gen_gamma} with the replacement $g\rightarrow 1/x$, and plugging it in Eq.~\eqref{ODE0}, one gets
\begin{align}\label{mainEqexplicit}
R'(x)&= -\frac{q\,R(x)}{\beta_1} -\frac{\beta_1(a+s)-\beta_2 q}{\beta_1^2x} R(x)   \nonumber \\
&- \left[ \frac{2q+r}{2\beta_1}+\frac{2\beta_1s-\beta_2(2q+r)}{2\beta_1^2x} \right] R^2(x)+\mathcal{O}(R(x))^3 \,,
\end{align}
or in terms of the anomalous dimension
\begin{align}\label{mainEqgamma}
\gamma(x)'=&-\frac{q\,\gamma (x)}{\beta_1} + \left[ \frac{a\beta_1(q+2r)-q(\beta_1s-\beta_2q)}{\beta_1^2qx} \right]\gamma(x) \nonumber \\
&- \left[ \frac{q+r}{\beta_1q} -\frac{a\beta_1r(3r-q)+2q(\beta_2q(q+r)+\beta_1rs)}{2\beta_1^2q^3x} \right]\gamma(x)^2+\mathcal{O}(\gamma^3) ,
\end{align}
where $\gamma_1 = a \frac{1}{x} + b \frac{1}{x^2}+... $. This equation is in the normal form presented in Ref.~\cite{Costin1995} and represents the basis upon which the isomorphism
of the generalized resummation is built (see~\ref{Appendix:nonliearODE}). Because of the nonlinearity of Eq.~\eqref{mainEqgamma}~\cite{Costin1995}(see the review in Ref~\cite{Dorigoni:2014hea}), the Borel transform of $\gamma$ has an infinite number of singularities at $q\,k/\beta_1$, $k\in\mathbb{N}^+$.
Therefore, identifying these with the UV renormalons we must set $q=1$, by matching for example with a direct calculation in the large number of fermions $N_f\rightarrow \infty$ limit in QED. Note that $q$ is a structural parameter relating the nonperturbative contribution of the Green function with that of the anomalous dimension, and thus it must be independent of any particular value of $N_f$.

\subsection{IR renormalons} \label{IRrenomalons}

So far we have focused on the UV renormalons, however, the above discussion also holds at the IR. The appearance of renormalon ambiguities is in any case linked to an energy scale~\cite{Cvetic:2018qxs}. Consider $\beta_1<0$: the minus sign can be reabsorbed
in the parameter $q$ when doing the identification with an explicit calculation in the literature, thus determining $q=-1$. We should point out that the Callan-Symanzik equation
is not homogeneous in the infrared region as in Eq.~\eqref{CS}. Nevertheless, this non-homogeneous term does not change the properties of the ODE and our conclusions still hold.

Apparently, in the IR case, the first singularity at $-1/\beta_1$ is missing, in contrast to the UV case. An explicit calculation of the skeleton diagram in the IR limit shows that the first renormalon is at $-2/\beta_1$ with the remaining poles spaced by $|-1/\beta_1|$. Conversely, within the resurgence framework,   IR and UV renormalons are indistinguishable. The fact that the first IR renormalon seems to be missing is not describable within the ODE and analyzable functions approach. This can be explicitly understood by noticing that, for non-linear systems, the infinite number of Borel poles are found recursively from the first singularity in the Borel plane. Thus the entire Stokes line is built from the first pole (see~\ref{Appendix:nonliearODE}).

Our interpretation of this apparent mismatch between explicit calculations of skeleton diagrams and general properties inferred by a specific ODE is as follows. Besides the skeleton diagrams, renormalon singularities are also found in Ref.~\cite{Parisi:1978iq} by transforming the Callan-Symanzik equation and at the same time using perturbation theory~\cite{Parisi:1978bj,Parisi:1978iq}. In particular, working on the 1PI renormalized $n-$point function $G^{(n)}_R(p)$, one gets singularities in the Borel plane at $(n-4)/(2\beta_1)$. Clearly, for $n=2$ and $\beta_1 < 0$, there is indeed one singularity at $1/|\beta_1|$ which has to be identified with the first IR renormalon coming from the analysis of ODE with $q=-1$.

\paragraph{Connections with asymptotically free gauge theories}
We have shown that within the framework of analyzable functions both abelian ($\beta_1>0$) and non-abelian ($\beta_1<0$) theories suffer the same renormalon singularities. Following Ref.~\cite{Parisi:1978iq}, in the case of non-abelian theory, one has to add a dimension two operator to reabsorb the singularity at $1/\beta_1$.
For a non-abelian gauge theory, the requirement of such an operator is not trivial, since one cannot add a mass term for the gauge boson because of gauge invariance. In Ref.~\cite{Dudal:2007rw}, it has been proposed that a gauge invariant,  dimension two nonlocal operator (still satisfying physical requirements of unitarity) can be added to cure the necessity of a dynamical generation of the gluon mass. Amusingly, a non-local operator  also emerges from the synthesis of the renormalons, i.e. resumming them in the generalized sense (at least in the $\phi^4$ model)~\cite{Maiezza:2019dht}. After this resummation is performed there is no need of adding the higher-dimensional operators of Ref.~\cite{Parisi:1978iq}.

\subsection{Higher order corrections and the resurgence formalism}
In this section we discuss how higher order corrections enter into the non-perturbative part of Green functions. As originally noted in Ref.~\cite{tHooft:1977xjm}, only the one-loop constant $\beta_1$ enters in all the non-perturbative corrections due to the renormalons. A natural question is whether higher order corrections ($\propto \beta_2,\beta_3,...$) may potentially modify these leading results -  and if so, how we can calculate these corrections systematically. We shall make use of the interplay between the synthesis and the analysis of the RGE to answer  these questions. In Ref.~\cite{Maiezza:2019dht} it was shown that any Green function may be written as (assuming the Green function is finite)
\begin{equation}
G(g)= \sum_{k=0}^{\infty}\,e^{-\frac{\,k\beta_1}{g}}\, \int_0^{\infty}dz\, e^{-\frac{\,z}{g}}(\mathcal{G}_k(z))_{bal} \,,\, \text{for } g < g_0\,,
\end{equation}
where $g_0$ is  real and positive and  the subscript $bal$ means that the balanced average of Ref.~\cite{costin1998} is taken. The function $\mathcal{G}_0(z)$ is the usual one obtained from perturbation theory and its analytic continuations. The $\mathcal{G}_k(z)$ functions with $k\geq1$ are disconnected functions, which are \emph{genuine} non-perturbative corrections coming from the analyzable functions theory.

Using the synthesis of the renormalons as in Ref~\cite{Maiezza:2019dht}, it is straightforward to show that only the first disconnected sector $\mathcal{G}_1(z)$ is non-null whenever the Borel transform for $\mathcal{G}_0(z)$ is of the form $\mathcal{G}_0(z)=1/(z_0-z)^n$, with $n$ an integer number (see Sec. 3 of Ref.~\cite{Maiezza:2019dht}). Note now from Eq.~\ref{mainEqgamma} that the higher-order corrections proportional to $\beta_2$ enter at the leading order in the second term proportional to $\gamma(x)/x$. In this case, the Borel transform of $\mathcal{G}_0(z)$ is of the form $1/(z_0-z)^{1+a}$, where the constant  $``a"$ is a real number (see~\ref{beyond_simple_poles} for details). Hence the non-perturbative sectors $\mathcal{G}_2(z), \mathcal{G}_3(z),...$ may be non-zero in this case.  Bottom line: all the corrections to the result of Ref.~\cite{Maiezza:2019dht} due to higher non-perturbative sectors must be proportional to $\beta_2$ (or higher terms of the beta function). An explicit realization of the latter point is presented in Ref.~\cite{Gardi:2001wg}.

\subsection{Remarks}

Some comments are now in order.

\begin{enumerate}

 \item The proof is generic in the sense that, starting only from RGE, together with resurgent analysis, we show the existence of renormalon ambiguities with no reference to the skeleton diagram~\cite{tHooft:1977xjm},
 or to perturbative and direct approaches on $n-$points correlation functions~\cite{Parisi:1978bj,Parisi:1978iq}. The identification with explicit calculations is only done \textit{a posteriori}.

 \item The skeleton diagram leads to poles in the Borel transform of any Green function~\cite{tHooft:1977xjm}. This is what has been recently resummed in a generalized sense in Ref.~\cite{Maiezza:2019dht}, leaving only a single undetermined parameter. In the framework of analyzable functions, the operation of resummation along a Stokes line must be in a one-to-one correspondence with an ODE, which in Ref.~\cite{Maiezza:2019dht} has been suggested to be the RGE.
   In the present work, we provide explicit proof of this particular point.

 \item The only technical assumption we have made is that the function $M(R,g)$ in Eq.~(\ref{gen_gamma}) is expandable in powers series of $R$.

 \item Starting from simple poles emerging from the direct estimation of simple skeleton diagrams, one ends up to simple poles plus branch point through Schwinger-Dyson equations~\cite{tHooft:1977xjm}, which is exactly what we get in the Borel transform of Eq.~\eqref{mainEqexplicit}. However, due to higher-order corrections, higher poles and branch points can emerge as well (see~\ref{beyond_simple_poles}).

\item We have set the parameter $q=1$ in Eq.~\eqref{mainEqexplicit} by \textit{a posteriori} matching the result with the known renormalon pole's position. Notice that there is no \textit{a priori} reason why $q=1$ from the analysis of the RGE. This invites us to imagine a hypothetical situation in which $q\ll 1$. In this case, the Stokes line would have an accumulation of singularities close to the origin. If so, one would be dealing with a situation in which the standard perturbation theory (also improved by standard BL resummation) would be completely ineffective and useless. In other words, the hypothetical theory under consideration would be perturbatively nonrenormalizable, and this suggests a possible link to gravity since it is indeed a  non-renormalizable theory from a perturbative approach.

\end{enumerate}

\section{Embedding in abelian gauge theory within Large $N_f$ expansion}\label{SEC:largeN}

In this section, we illustrate in some specific cases how the non-perturbative corrections propagate to the beta function\footnote{Notice that, although it is a non-perturbative object in general, only the asymptotic expansion of $\beta(g)$ is relevant for the emergence of renormalons from RGE.}.
In particular, we expand our discussion within the specific frameworks of QED and QCD with a large number of fermions ($N_f$) and we show how renormalons appear in the beta function of the models. The clear advantage is that the beta function is known to all orders in perturbation theory and leading order in $1/N_f$. This fact enables us to better illustrate some concepts above elaborated. Furthermore, instantons singularities are suppressed in the large $N_f$ limit~\cite{PhysRevD.15.1655,KOPLIK1977109}, thus making this framework ideal for studying renormalon properties.

\paragraph{Renormalons and renormalization schemes}
We revisit the appearance of renormalons singularities in the large $N_f$ limit and within different renormalization schemes. While apparently there are no renormalons for example in the $\beta_{\overline{MS}}$, on the contrary, there are in the beta function in the MOM-scheme $\beta_{MOM}$. Does this scheme dependence means that the renormalon singularities are unphysical? The answer to this question was given several years ago in Refs.~\cite{Grunberg:1980ja,Grunberg:1982fw,BENEKE1993154}, but it might be useful to the reader going briefly through it for completeness.

In Ref.~\cite{BENEKE1993154}, it was concluded that one cannot enforce consistency on the asymptotic expansion in any
arbitrary renormalization scheme. This is why an effective charge beta-function was introduced in Ref.~\cite{Grunberg:1980ja,Grunberg:1982fw}, to avoid the choice of renormalization scheme at all orders (that is indeed mathematically ill-defined for an asymptotic series). The effective
charge beta-function makes use only of the non-perturbatively well-defined quantity $D(Q^2)$, i.e. the Adler function that we further discuss below. The physicality of renormalons thus has to be regarded beyond perturbation theory and hence independently of any renormalization scheme.

From the above discussion, it should be clear to the reader that the renormalon singularities may manifestly be in the beta function in some schemes and not in others, as it is illustrated below with examples. The bottom line is that, when they are not manifest, there should be a way in which they emerge in the beta functions. For example, in the $\overline{MS}$-scheme,  although the $n!$-growth does not naively appear in the beta functions, this factorial growth still appears in the finite part of Green functions (as the original case presented in Ref.~\cite{tHooft:1977xjm}). These ambiguities due to the $n!$-growth in the Green functions  may  be reabsorbed using  higher-dimensional operators $\phi(x)^m$ with $m\geq 6$ as counterterms~\cite{Parisi:1978iq}~\footnote{ In Ref.~\cite{Parisi:1978iq} this procedure has been  illustrated for the $\phi^4$ model. For a discussion in the context of gauge theories see Ref.~\cite{Altarelli:1994vz,Ball:1995ni} (see also Ref.~\cite{Altarelli:1995kz} for a review, and the reference therein). Let us recall once again, that an improved approach with respect to the one provided by Parisi in Ref.~\cite{Parisi:1978iq}, is the generalized resummation in Ref.~\cite{Maiezza:2019dht}, which ultimately leads to an effective non-local operator.}, which are exponentially suppressed for small coupling constant $\lambda$ as $e^{-\frac{2n}{\beta_1\lambda}}$.
These higher-dimensional counterterms lead to nonanalytic contributions to the running of all the couplings in a theory (see e.g. Ref.~\cite{Wilson:1970ag}). Finally, these non-analytic contributions are the source of the singularities in the Borel transform.

In what follows, we shall discuss in more detail the $\overline{MS}$-scheme, the  on-shell scheme and in particular the effective charge beta-function.

\subsection{The beta function in the minimal subtraction scheme}

In the minimal subtraction ($\overline{MS}$) scheme, at the leading non-trivial order in the $1/N_f$ expansion, the QED beta-function reads~\cite{palanques-mestre1984,Gracey:1996he} (see Ref.~\cite{Holdom:2010qs} for a recap)
\begin{equation}
 \beta^{QED}_{\overline{MS}} (A)=\frac{2 A^2}{3}+\frac{A^2}{2 N_f}\int_0^A dx F_{QED}(x)+\mathcal{O}\left(\frac{1}{{N_f^2}}\right)\,,
\end{equation}
where
\begin{equation}
 F_{QED}(x)=-\frac{(x+3)(x-\tfrac{9}{2})(x-\tfrac{3}{2}) \text{sin}(\tfrac{\pi x}{3})\Gamma(\tfrac{5}{2}-\tfrac{x}{3})}{27 \cdot 2^{\tfrac{2x}{3}-5} \pi^{\tfrac{3}{2}}(x-3)x \Gamma(3-\tfrac{x}{3})}\,.
\end{equation}
An apparent feature of this renormalization scheme is that the perturbative series for $ \beta_{\overline{MS}} (A)$ is renormalon free, as discussed in Ref.~\cite{Antipin:2018asc}.

As already discussed in the introduction of this section, this is not the end of the story; in fact, in the $\overline{MS}$ scheme one absorbs the infinite part of any computation at a given \emph{finite} order in the coupling expansion in the counterterms. Whereby, the finite part in the Green functions remains. This finite part goes as $n!$ (at order $A^n$) and then it makes the expressions for Green functions non-Borel-summable, then ambiguous. To have well-defined expressions, these ambiguities
may be reabsorbed into higher dimensional operators~\cite{Parisi:1978bj,Parisi:1978iq,Altarelli:1994vz,Ball:1995ni} which, in turn, change the RGE running of the coupling $A$. Note that this is a general feature of the strongly interacting QFTs (in the sense of nonperturbatively treatable)~\cite{Wilson:1970ag,Wilson:1973jj}.

In fact, similar discussion holds for non-abelian gauge theories. For example, also in QCD there are UV renormalon singularities at the position $2k/\beta_1$ with $k=1,2,3,...$. However, since these poles are in the negative real  Borel axis (for QCD $\beta_1<0$), they do not cause ambiguities when computing the Laplace transform for UV quantities. More
important, there are IR renormalons~\cite{tHooft:1977xjm}, which are singularities at $-2k/\beta_1$ with $k=2,3,4,...$ in the Borel transform of the Green functions. These singularities are important when computing low energy quantities and are related with the problem of the strong coupling in the IR regime of QCD -- see also Sec.~\ref{IRrenomalons}.

Although this clear physical manifestation of the renormalons in low scale QCD~\cite{Bauer:2011ws}, the $\overline{MS}$ beta function at the leading order in the $1/N_f$ expansion is a well defined function, in full analogy with the above discussed QED case. In particular, for $SU(N_C)$ at the leading order in the $1/N_f$ expansion, the beta function reads as~\cite{Gracey:1996he}
\begin{equation}
 \beta^{QCD}_{\overline{MS}} (A)=\frac{2 A^2}{3}\left(1-\frac{11 C_2(G)}{4 N_f S_2(R)}\right)+\frac{A^2}{2 N_f}\int_0^A dx F_{QCD}(x)+\mathcal{O}\left(\frac{1}{{N_f^2}}\right)\,,
\end{equation}
where
\begin{align}
F_{QCD}(x)=&\frac{2^{1-\tfrac{2x}{3}}\text{sin}(\tfrac{\pi x}{3}) \Gamma(\tfrac{5}{2}-\tfrac{x}{3})}{27 \pi^{3/2}(x-3)^2 x \Gamma(3-\tfrac{x}{3})}\left[\frac{C_2(G)}{S_2(R)}(4x^4-42x^3+288x^2-1161x)\right. \nonumber \\  & \left. -4\frac{d(G)}{d(R)}(x-3)(x+3)(2x-9)(2x-3) \right] \ \ .
\end{align}
$S_2(R)$ and $C_2(G)$ are quadratic Casimirs and $d(R)$, $d(G)$ are, respectively, the dimensions of the representation $R$ and of the group $G$.

Bottom line: both for the abelian and non-abelian cases, the perturbative series for $\beta_{\overline{MS}} (A)$ are renormalon free and they can be resummed in $F_{QED,QCD}(x)$. This property seems a characteristic feature of the $\overline{MS}$ scheme and is not related to asymptotic freedom.

\subsection{The beta function in the on-shell scheme}

The beta function for QED in the on-shell scheme reads~\cite{Broadhurst:1992si}
\begin{equation} \label{BoB}
 \beta^{OS}(A)=\frac{2}{3}A^2+\frac{1}{N_f}\sum_{n=2}^{\infty}\beta_n^{OS}\,A^n+\mathcal{O}\left(\frac{1}{N_f^2}\right)\,,
\end{equation}
where
\begin{equation} \label{betaOS}
\beta_n^{OS}=\frac{1}{2}\delta_{n,2}-\frac{7}{9}\delta_{n,3}+\frac{n-2}{3}\int_0^1 d\theta(N_1(\theta)+N_2(\theta))\frac{d}{d\theta}[-w(\theta)]^{(n-3)}
\end{equation}
and
\begin{align} \label{wditheta}
 & N_1(\theta)=-\frac{2\theta^2(3+2\theta+3\theta^2)\log(\theta)}{(1-\theta^2)^3}-\frac{\theta(1+6\theta+\theta^2)}{(1-\theta^2)^2}  \\
 & N_2(\theta)=\frac{2\theta^2(3-\theta)\log(\theta)}{(1-\theta)^3}+\frac{\theta(28-9\theta+6\theta^2-\theta^3)}{6(1-\theta)^2} \\
 & w(\theta)=\frac{5}{9}-\frac{4\theta(1-\theta)-(1-4\theta+\theta^2)(1+\theta)\log(\theta)}{3(1-\theta)^3}\,. \label{wditheta}
\end{align}
Here $\theta$ is an integration variable defined by: $-q^2=m^2 \tfrac{(1-\theta)^2}{\theta}$, where $q$ and $m$ stay for the mass and the momentum flowing in a fermion bubble, respectively.

One can isolate the renormalon singularities by performing the UV expansion ($q^2 \gg m^2$). To this end, we resort to the Newton method to invert the transcendental function $w(\theta)$, choosing as initial guess
\begin{equation}
\theta = e^{3w-5/3}\,,
\end{equation}
which is the inverse of $w(\theta)$ in the UV limit $\theta \to 0$ \footnote{This choice substantially increases the convergence of the method around $\theta = 0$.}. Then, we expand $N_{1,2}$ in powers of $w$ and $e^{3w-5/3}$ and we solve the integral in \eqref{betaOS} using $w$ as integration variable. This gives
\begin{equation} \label{order}
  \frac{\beta_n^{OS}}{(n-2)!}=\frac{1}{2}\delta_{n,2}-\frac{7}{9}\delta_{n,3}+\Theta(n-4)\sum_{k=1}^{\infty} \frac{P_k(n)}{\left(\tfrac{2}{\beta_1}k\right)^n e^{\tfrac{5 k}{3}}} \quad\,.
\end{equation}
$\Theta(n-4)$ is the Heaviside step function and $P_k$ are polynomials in $n$ of degree $J$ where
\begin{equation}
  J=\begin{cases}
  \frac{k-2}{2} \ \ \text{if} \ \ k \ \ \text{is} \ \ \text{even} \\ \frac{k-1}{2} \ \ \text{if} \ \ k \ \ \text{is} \ \ \text{odd}
\end{cases} \ \,.
\end{equation}
Multiplying Eq. \eqref{order} by $t^{n-2}$ and summing over $n$ we get the Borel transform of the $1/N_f$ beta function \footnote{Before to compute the Borel transform, we brought a factor of $A$ out of the sum in Eq.~\eqref{BoB}.}
\begin{equation}\label{ostrasform}
\mathcal{B}\left[\sum_{n=2}^{\infty}\beta_n^{OS}\,A^{n-1}\right](t)=\frac{1}{2}-\frac{7}{9}t+\sum_{k=1}^{\infty}t^2 e^{-\tfrac{5 }{3}k}\frac{B_k(t)}{(t-\tfrac{2}{\beta_1}k)^{J+1} }
\end{equation}
where $t$ is the transformed coupling and $B_k(t)$ are polynomials of degree $J$ which are regulars in $t=2 k/\beta_1$. The polynomials $P_k(n)$ and $B_k(t)$ are listed in~\ref{App:OSBETA} up to $k=8$.
Clearly, the above expression contains singularities on the real positive axis at $2 k/\beta_1$, with $\beta_1=\tfrac{2}{3}$. These are renormalon poles of increasing order for increasing values of $J$.

\section{Adler function and effective charge in QED}\label{SEC:cha}

In this section, we finally come back on the effective charge beta function mentioned in the introduction of Sec.~\ref{SEC:largeN}. It is a neat way of illustrating our general result on the renormalons in the beta function presented in Sec.~\ref{SEC:MAIN}.

The scheme-independent coupling in QED is defined by
\begin{equation}
\alpha (Q^2) = \frac{\alpha(0)}{1-[\Pi(Q^2)-\Pi(0)]}\,,
\end{equation}
where $\Pi(Q^2)$ is the self energy of the photon field. Taking the derivative with respect to the scale $Q^2 \equiv \mu^2$, one has
\begin{equation}
\beta^{\text{Adler}}(\alpha)= \frac{d\alpha(\mu^2)}{d\mu^2} = \frac{1}{4\pi^2} \frac{\alpha(\mu^2)}{\alpha(0)} D(\mu^2)\,,
\end{equation}
where $D$ is the so-called Adler function $D(\mu^2)= 4\pi^2 d\Pi(\mu^2)/d\mu^2$. This function is known to be affected by renormalons and its precise form can be found for instance in Ref.~\cite{Beneke:1998ui}. In the QED case, the Adler function exhibits simple and quadratic poles in the positive real axis of its Borel transform.

\section{Outlook}

The present work has explored some issues of the classical Borel-Laplace approach in connection with the renormalization group. In particular, we have focused on the singularities along the integration line of the Laplace integral known as renormalons. These singularities produce ambiguities in the Borel-Laplace method, potentially destroying the paradigm of reconstructing a Green function from its perturbative expression. While a consistent solution has been recently proposed in Ref.~\cite{Maiezza:2019dht},
in this article we have analyzed the formal emergence of renormalons from the renormalization group equation. Specifically, we have proved that the Borel transform of a given Green function satisfying the renormalization group equation has a Stokes line with poles spaced by $1/|\beta_1|$.

The proof has exploited the framework of ODEs and analyzable functions~\cite{Costin1995}, starting only from the observation that the two-point function contains a non-perturbative finite part. If this were not the case, then Eq.~\eqref{BLrepr}  would  be enough  to reconstruct the two-point correlation function via the Laplace integral. Notice  that  in our analysis there is no \textit{a priori} reference to any specific set of Feynman diagrams~\cite{tHooft:1977xjm}, or any higher-dimensional  counterterms~\cite{Parisi:1978bj,Parisi:1978iq}. It is worth stressing that our result makes manifest the robustness of the resummation of renormalons based on a one-parameter transseries of Ref.~\cite{Maiezza:2019dht}.

After a careful discussion on the validity of our argument and on the connection with the previous results known in the literature, we have considered gauge theories in the limit of a large number of fermions. It is well known that this expansion provides a limit scenario in which instantons, another possible source of ambiguities in the Laplace transform, are naturally suppressed while leaving unchanged the renormalon poles. We aimed to make as manifest as possible the appearance of non-perturbative effects through RGE and analyzable functions.

Finally recall that, since the renormalon's nature is non-perturbative, the position of the poles in the Borel transform is independent of any renormalization scheme. The specific choice of the on-shell scheme is particularly  useful in order to show how the $n!$ contributions enter in the $\beta-$function, but it is not by any mean the only way of showing their emergence. The bottomline is that  these non-perturbative corrections allow exploring the possibility that the UV fate of a theory may be driven by non-perturbative physics. In turn, the requirement of the finiteness  of the theory (absence of Landau poles) at any energy might alleviate the issue of non-uniqueness of the generalized resummation of renormalons.

\section*{Acknowledgement}

We thank Oleg Antipin for countless discussions and Gorazd Cveti\v{c} for a careful reading of the manuscript and useful comments.
JB and AM were partially supported by the Croatian Science Foundation project number 4418. JCV was supported in part under the U.S. Department of Energy contract DE-SC0015376.

\appendix

\section{Recursive relations from the renormalization group equation }\label{rec}

A recursive equation for the coefficient $\gamma_i$ of Eq.~\eqref{G_PT2} can be extracted from RGE.~\cite{Yeats:2008zy,Klaczynski:2013fca}. This still holds here for Eq.~\eqref{G_true}, but the difference is that now the recursion starts to $\gamma_1=\gamma-M(R)\approx \gamma-R$. As a consequence, the nonperturbative piece $R$ propagates throughout the recursion. One gets the $\gamma$s by replacing Eq.~\eqref{G_true} in Eq.~\eqref{CS} and solving at each power of $L$. The expression is given by
\begin{equation}
- \beta(g) \gamma _k'(g)+\gamma(g) \gamma _k(g)+(k+1) \gamma _{k+1}(g) =0\,,
\end{equation}
for example when $k=1$, then
\begin{equation}
\gamma_2 \simeq \frac{1}{2} \left(\beta(g) \gamma _1'-\gamma _1^2 - \gamma _1R \right)\,,
\end{equation}
and so on, exactly as in Ref~\cite{Klaczynski:2013fca} except that there is now the correction due to $R$ to every $\gamma$s. It is manifest that $R$ is the only
term ruling the nonperturbative part of the correlation function.

\section{On the non-linearity of ODE}\label{Appendix:nonliearODE}

The fundamental normal ODE for this work (coming from a double-expansion of a generic ODE for small $y$ and $1/x$) is written as~\cite{Costin2008}
\begin{equation}\label{generic}
y'(x)=f(x_0)-\lambda y+a/x y+g(x,y)  \,,
\end{equation}
with $g(x,y)=\mathcal{O}(x^{-2}|y^2|x^{-2}y)$. In Sec.~\ref{SEC:MAIN} we have identified this equation with our
Eq.~\eqref{mainEqexplicit} and stated that its Borel transform stems infinite singularities in $n \lambda$, with $n\in \mathbb{N}^+$.
The reason is the non-linearity of $y$ in $g(x,y)$. For the sake of completeness, in this appendix we want to directly illustrate this.

\subsection{A simple case}
Consider for example the equation
\begin{equation}\label{example}
y'=-\lambda y+y^2+1/x  \,,
\end{equation}
and Borel transform (convolutive model)
\begin{equation}
-z Y=-\lambda\, Y +Y*Y+1  \,,
\end{equation}
where the star denotes the convolution. So we have
\begin{equation}
Y=-\frac{1}{z-\lambda} -\frac{Y*Y}{z-\lambda} \,,
\end{equation}
and introducing a auxiliary parameter $\epsilon$ on the convolutive product for an iterative expansion, leads to
\begin{equation}
Y=-\frac{1}{z-\lambda} -\epsilon\frac{Y*Y}{z-\lambda} \,.
\end{equation}
When $\epsilon=0$, $Y=-\frac{1}{z-\lambda}$. On this initial solution is then possible to iterate in power of $\epsilon$:
\begin{align}\label{step1}
Y=& -\frac{1}{z-\lambda} -\epsilon\frac{1}{z-\lambda} \int_0^z \frac{1}{z_1-\lambda}\frac{1}{z-z_1-\lambda}dz_1 = \\ \nonumber
&= -\left( \frac{1}{z-\lambda} + \epsilon \frac{2\log(\lambda-z)}{(z-\lambda ) (z-2 \lambda )} \right) \,,
\end{align}
which present simple poles in $z=\lambda,2\lambda$ and a branch point in $z=\lambda$. Going on with the iteration from Eq.~\eqref{step1} to
higher powers in $\epsilon$, one covers at all orders the infinite set of poles and the statement of Sec.~\ref{SEC:MAIN} results proved.

\subsection{An example from the Callan-Symanzik equation}

For the $\phi^4$ model, one must expand more in Eq.~\eqref{mainEqgamma} and keep terms up to order $1/x^2$, since $\gamma_1\propto 1/x^2+\mathcal{O}(1/x^3)$ $(g=1/x)$. Neglecting the coefficients $\beta_2$ and $s$ (the latter coming from mixed products $R\times 1/x$), and finally choosing some constants arbitrarily to exemplify, we get
\begin{equation}\label{example2}
y'=-y+y/x^2-y^2(1+1/x^2)+1/x^2 \,,
\end{equation}
where we have denoted in this sketch equation the unknown function with $y(x)$. The aim here is to consider an equation slightly more complicated than Eq.\eqref{example} and even with $1/x^2$ suppression terms.

The Borel transform of $y/x^2$ is
\begin{equation}
\mathcal{B}[y/x^2]= \int_0^z z_1 Y(z-z_1)dz_1 \,,
\end{equation}
so the Borel transformed Eq.\eqref{example2} is
\begin{equation}
Y(z)= \frac{1}{1-z}\left[\int_0^z z_1 Y(z-z_1)dz_1- \epsilon \left(\delta(z)+z\right)*Y(z)*Y(z) +z \right]\,,
\end{equation}
where we put the auxiliary parameter $\epsilon$ for the expansion, exactly as in the case discussed above in the previous subsection.
We start again with $\epsilon=0$ and, taking the second derivative of this expression, we get
\begin{equation}
Y''(z-1)+2 Y'+Y=0\,,
\end{equation}
leading to the formal solution
\begin{equation}\label{formalB}
Y_0(z)=\frac{c_1 I_1\left(2 \sqrt{1 -z}\right)}{\sqrt{1 -z}}-\frac{c_2 K_1\left(2 \sqrt{1 -z}\right)}{\sqrt{1 -z}}\,,
\end{equation}
where $I_1,K_1$ are the modified Bessel function of first and second kind respectively, $c_{1,2}$ arbitrary constants and $Y_0(z)$ is the same defined in Ref.~\cite{Costin1995}. The function $I_1$ is holomorphic and $K_1$ is singular at $z=1$. Expanding around $z=1$ we obtain
\begin{equation}\label{sing_expand}
Y_0(z)=-\frac{c_2}{2} \left(\frac{1}{z-1}+\log (z-1)\right)+\text{const.} +\mathcal{O}(\sqrt{z-1}) \,.
\end{equation}
Thus, the same conclusion of the example in the first paragraph of this appendix holds:
replacing iteratively this expression in the quadratic term $Y*Y$, one gets a series in $\epsilon$ showing infinite poles
spaced by unit. It is finally worth recalling that a solution of Eq.~\eqref{example2} can be written as a simple transseries~\cite{Costin1995} - if one searches a solution of Eq.~\eqref{example2} in terms of a formal power series, the generalized resummation on such series gives the transseries solution.

\subsection{Higher poles}\label{beyond_simple_poles}

In this subsection we recall the connection between the ODE and the type of singularity in the Borel transform of its solution. Consider the following equation
\begin{equation}
y'=-\lambda y+y^2+a \frac{1}{x} y + f_0(x) \,,
\end{equation}
where the $f_0(z) $ denotes possible higher order terms whose precise form is not relevant for the discussion to follow. Notice that in Eq.~\eqref{mainEqgamma}, there is a term $\propto \frac{1}{x} \gamma$ corresponding to the ``a" term of the equation above.

Applying the Borel transform we have
\begin{equation}
-zB(z) = -\lambda B(z) +a  \int_0^{z} d s B(s) + B_0(z) \, ,
\end{equation}
or
\begin{equation}
(\lambda-z)B(z) = a  \int_0^{z} d s B(s) +B_0(z)\, ,
\end{equation}
where $B_0$ is just the Borel transform of $f_0$. Taking the derivative of the above equation, the solution to this integral equation for $a\neq -1$ is given by (up to an arbitrary constant),
\begin{equation}
B(z) = \frac{c}{(\lambda-z)^{1+a}} - \frac{1}{(\lambda-z)^{1+a}} \int_0^z B_0(s)(s-\lambda)^{a} ds \,.
\end{equation}
When $a =-1$, $B(z)\propto \log (z-\lambda)$. Thus, we see that the constant $a$ is related to the type of singularity in the Borel transform of the solution to the ODE under consideration. This equation together with Eq.~\ref{mainEqgamma} shows how higher order loop corrections modify the type of singularities of the Borel transform. It is instructive to compare it with explicit evaluation in QCD observables where several types of singularities are found~\cite{Cvetic:2018qxs}.

\section{On-shell QED beta function: polynomials $P_k(n)$ and $B_k(t)$}\label{App:OSBETA}

For the sake of completeness, we provide some explicit results for the renormalon expansion of the large $N_f$ QED beta function in the on-shell scheme.

The polynomials $P_k(n)$ appear in Eq. \eqref{order}. For $k \le 8$, their explicit form is
 \begin{align}
   & P_1(n)=-33 \qquad \qquad P_2(n)=-1188 \qquad \qquad P_3(n)=-81(355-34n)\\ \nonumber \\ \nonumber
   & P_4(n)=-192(2999-437n) \qquad \qquad
   P_5(n)=-15(679463+6n(-20035+273n)) \\ \nonumber \\ \nonumber & P_6(n)=-\frac{648}{5}(1288251+5n(-51392+1493n)) \\ \nonumber \\ \nonumber
   & P_7(n)=-\frac{7}{5}(1854295183-30n(13402844+n(-578339+2358n))) \\ \nonumber \\ \nonumber & P_8(n)=-\frac{384}{35}(3544414470-7n(116808583+15n(-431132+4245n))) \qquad .
 \end{align}
The first eight polynomials $B_k(t)$ of Eq. \eqref{ostrasform} are
 \begin{align}
  &  B_1(t)=\frac{11}{9} \qquad \qquad
   B_2(t)=\frac{11}{2} \qquad \qquad
   B_3(t)=\frac{253 \ t-1971}{9} \qquad \qquad \\ \nonumber \\ \nonumber
  & B_4(t)=\frac{4}{9}(422 \ t-3753) \qquad \qquad
   B_5(t)=\frac{2023479+t(-333078+13343 \ t)}{9} \qquad \\ \nonumber \\ \nonumber
  & B_6(t)=\frac{2}{15}(20511954+t(-2848431+97426 \ t))\qquad \qquad \\ \nonumber \\ \nonumber
  &
   B_7(t)=\frac{-98096530707+t(16266030471+t(-895784481+16373077 \ t))}{135} \qquad \qquad \\ \nonumber \\ \nonumber
  &
   B_8(t)=\frac{8}{105}(-139615128864+t(20076468504+t(-965168405+15487702 \ t))) \qquad .
 \end{align}

\bibliographystyle{jhep}
\bibliography{biblio}

\end{document}